\documentclass[12pt]{iopart}

\usepackage{color}
\usepackage{graphicx}
\usepackage{caption}
\usepackage{rotating}

\begin{document}
%\TDM

\title{Structure of twisted and buckled bilayer graphene}

\author{Sandeep K. Jain}
\address
{Institute for Theoretical Physics, Universiteit Utrecht,
Princetonplein 5, 3584 CC Utrecht, The Netherlands}
\ead{S.K.Jain@uu.nl}

\author{Vladimir Juri\v ci\' c}
\address
{Nordita, Center for Quantum Materials, KTH Royal Institute of Technology and
Stockholm University, Roslagstullsbacken 23, S-106 91 Stockholm, Sweden}

\author{Gerard T. Barkema}
\address
{Department of Information and Computing Science, Universiteit Utrecht,
Princetonplein 5, 3584 CC Utrecht, The Netherlands}

\begin{abstract}
We study the atomic structure of twisted bilayer graphene, with very
small mismatch angles ($\theta \sim 0.28^0$), a topic of intense recent
interest. We use simulations, in which we combine a recently presented
semi-empirical potential for single-layer graphene, with a new term for
out-of-plane deformations, [Jain et al., J. Phys. Chem. C, 119, 2015] and an often-used
interlayer potential [Kolmogorov et al., Phys. Rev. B, 71, 2005]. This combination of
potentials is computationally cheap but accurate and precise at the same
time, allowing us to study very large samples, which is necessary to reach
very small mismatch angles in periodic samples.  By performing large scale
atomistic simulations, we show that the vortices appearing in the Moir\'e
pattern in the twisted bilayer graphene samples converge to a constant
size in the thermodynamic limit. Furthermore, the well known sinusoidal
behavior of energy no longer persists once the misorientation angle
becomes very small ($\theta<1^0$). We also show that there is a significant
buckling after the relaxation in the samples, with the buckling height
proportional to the system size. These structural properties have direct
consequences on the electronic and optical properties of bilayer graphene.
\end{abstract}
\maketitle

\section{INTRODUCTION}

Bilayer graphene (BLG) consists of two stacked graphene sheets with
usual stacking of either AB (Bernal) or AA type. However, two graphene
layers can also be placed on top of each other in other arrangements,
characterized in general by a mismatch angle $\theta$. Such
a structure is usually referred to as twisted bilayer graphene
(TBLG)\cite{Ohta2006,Reina2009}, and represents an example of a Van der
Waals heterostructure \cite{Geim2013}. Since TBLG is made of two stacked
misaligned lattices, a superlattice with a larger periodicity known as
Moir\' e pattern emerges in the structure \cite{Lu2013,Woods2014}.

Recently, this form of BLG has attracted a lot of attention
theoretically and experimentally due to its exotic electronic
\cite{Mele2010,Santos2007,Shallcross2008,Gail2011,Bistritzer2011,Morell2011,Chae2012,Pelc2015}
and optical properties \cite{Moon2013,Tabert2013,DeCorato2014}
arising due to the formation of the Moir\' e patterns. In particular,
it has been theoretically suggested that the twist in the BLG may
lead to a renormalization of the Fermi velocity \cite{Yin2015},
possible appearance of the flat electronic bands \cite{Morell2010},
neutrino-like oscillation of Dirac fermions \cite{Xian2013} as well as
localization of electrons \cite{Laissardiere2010}. Moreover, TBLG when
placed in a magnetic field, exhibits a fractal spectrum of the Landau
levels \cite{Wang2012}. This theoretical interest has been motivated
by the experimental observation of TBLG with Moir\' e patterns in the
samples grown on SiC substrates\cite{Hass2008}, and using chemical vapour
deposition \cite{Luican2011,Liu2015}. Furthermore, the mismatch angle
has a significant impact on the quantum Hall effect in TBLG, as has been
recently reported\cite{Lee2011}; and breaking of the interlayer coherence
for very small angles was experimentally found as well \cite{Kim2013}.

Out-of-plane buckling has a long-range effect in monolayer graphene and has significant impact on its structural properties and defect mechanics \cite{Fasolino2007,Katsnelson2008,Ma2009,Chen2011,Sandeep-2015}. The resulting nanometer sized ripples have been studied experimentally by transmission electron microscopy (TEM) \cite{Meyer2007} and scanning probe microscopy \cite{Ishigami2007,Stolyarova2007,Capasso2013,Ackerman2016}. Recently, out-of-plane ripples in bilayer graphene have been detected and investigated via TEM \cite{Butz2014,Meyer2007a} and the combination of dark-field TEM with scanning transmission electron microscopy (STEM) \cite{Lin2013}. The buckling effect in bilayer graphene has also been studied using computer simulations \cite{Wijk2015,Dai2016,Dai2016a}.

From the point of view of its electronic and structural properties, TBLG
is most interesting when the mismatch angle is small.  To theoretically
obtain the structure of the TBLG with periodic boundary conditions and
such small mismatch angles, it is necessary to consider samples with a
large size. An approach based on an effective elastic potential is quite
appealing in this regard since it allows to treat systems containing
millions of atoms.  Since we do not restrict ourselves to completely
flat graphene layers but allow for some buckling, we use a combination of
potentials, based on the recently developed semi-empirical potential for
the monolayer graphene \cite{Sandeep-2015} with a new term describing
out-of-plane deformations, and the more standard registry-dependent
interlayer graphitic potential \cite{KC-2005} to simulate relaxed
large bilayer graphene structures with very small mismatch angle
($\theta\approx0.28^{\circ}$). This combination of potentials is
computationally cheap and accurate.

In energy-minimized (relaxed) samples, vortices arise in the atomic
displacement field, due to the energy differences between different
kinds of stacking, in agreement with previous studies \cite{Wijk2015,Dai2016a,Uchida2014}. We show that after relaxation the size of these vortices
approaches a constant in the thermodynamic limit. Furthermore, we study
out-of-plane buckling in bilayer graphene and find that the buckling
height increases linearly with system size. We show that the buckling in
the pristine bilayer graphene is significant ($\sim3$ \AA~) and forms
a Moir\'e pattern analogous to the in-plane displacement without any
singularities and with long range structural effects.

\section{METHOD}

We use a new combination of intra-layer and inter-layer potentials
to simulate bilayer graphene. For the interactions within
the same layer, we use the recently developed semi-empirical potential
for single-layer graphene by Jain et al. \cite{Sandeep-2015},
which has a new out-of-plane deformation term.
The interlayer interactions are defined
by the registry-dependent Kolmogorov-Crespi potential without the local
normals \cite{KC-2005}. This combination of empirical potentials is
precise and accurate enough to capture the physical and structural
changes in the system without any heavy computational requirements. These
properties of the potential gives us freedom to study very large samples,
which is required for having very small mismatch angles under periodic
boundary conditions. In all samples studied, the energy is locally minimized,
starting from well-informed choices for the initial configurations:
insight obtained from many simulations of small systems is exploited
to start the energy minimization of large samples from already well-relaxed samples.
In our samples, we define a local energy per atom as follows: contributions
due to two-body interactions are equally divided over the two interacting
atoms, and contributions due to the three-body (angular) interactions are
attributed to the central atom. Thus, the sum of the local energy over all
atoms equals the total energy. This definition of local energy helps us to
visualize the local degree of mechanical relaxation in the sample.

\section{RESULTS AND DISCUSSION}

\begin{figure}
\begin{center}
\includegraphics[width=1.0\textwidth]{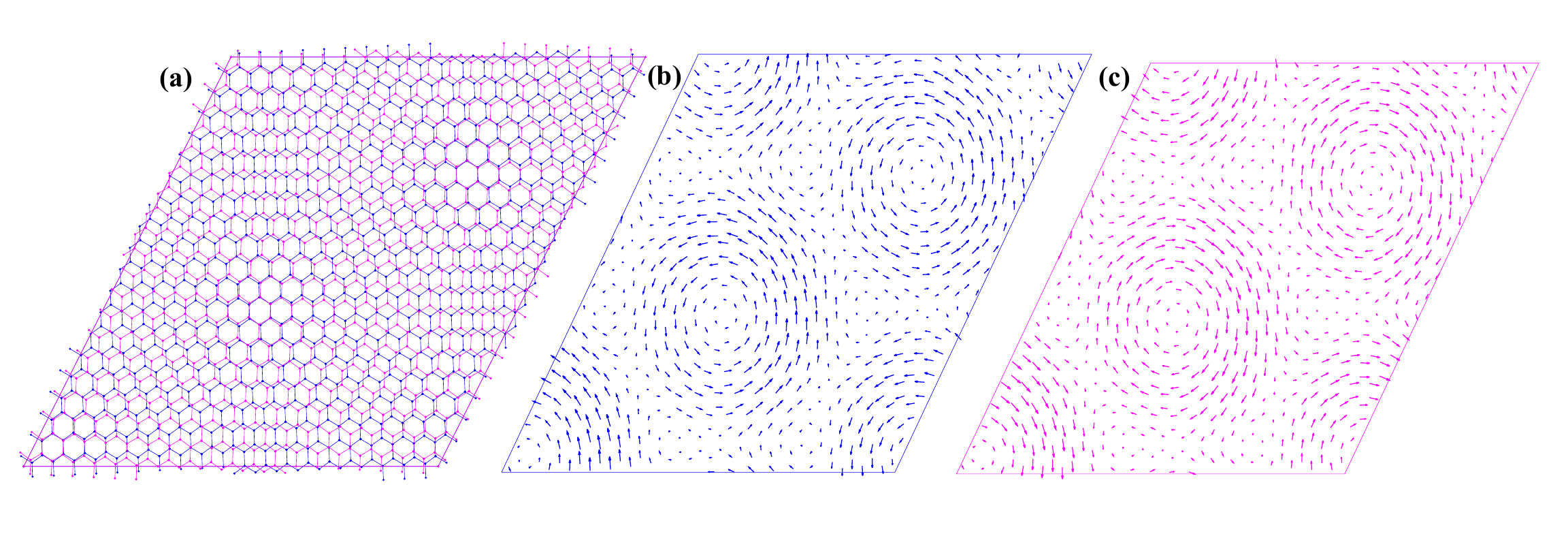}
\caption{Bilayer graphene sample with 1524 atoms and with mismatch
angle $\theta=5.09^{\circ}$. (a) Sample before relaxation. (b) Effect
of relaxation in the bottom layer where atoms rotate counterclockwise
near the vortex to minimize the total energy. (c) Effect of relaxation
in the top layer where atoms rotate clockwise near the vortex in order
to minimize the total energy. For visibility, the displacement arrows are enlarged by a factor of 20. }
\label{figure_1}
\end{center}
\end{figure}

We start with a sample having 1524 atoms in both layers with
a mismatch angle of $\theta=5.09^{\circ}$ between the layers as shown in
Fig.\ref{figure_1}(a). The Moir\' e patterns are clearly visible
in the sample along the diagonal. This sample is relaxed with the
above described combination of potentials, and its effect on atomic
relaxation in bottom and top layers is shown in Fig.\ref{figure_1}(b)
and Fig.\ref{figure_1}(c), respectively. The arrows in the figure
describe the relative atomic displacement after the relaxation with
respect to the unrelaxed positions (i.e. the positions in top and bottom
layer in the crystalline state of the individual graphene layers).
Atoms near the center of AA stacking
rotate to minimize the total energy and show a Moir\' e pattern of
displacement vectors with respect to their initial positions in the
form of vortices. In this case atoms in the bottom layer rotate in the
counterclockwise direction whereas atoms in top layer rotate clockwise,
since the center of mass of the system is unaltered. During the process
of relaxation the AA-stacked area becomes smaller while AB-stacked area grows,
since the energy of AB-stacking is lower compared to AA-stacking. This
result is in agreement with previous studies on TBLG \cite{Wijk2015,Uchida2014}. Relaxed bilayers have the intrinsic ripples in the structure
and the equilibrium separation between the layers is 3.46 \AA~.

\begin{figure}
\begin{center}
\includegraphics[width=1.0\textwidth]{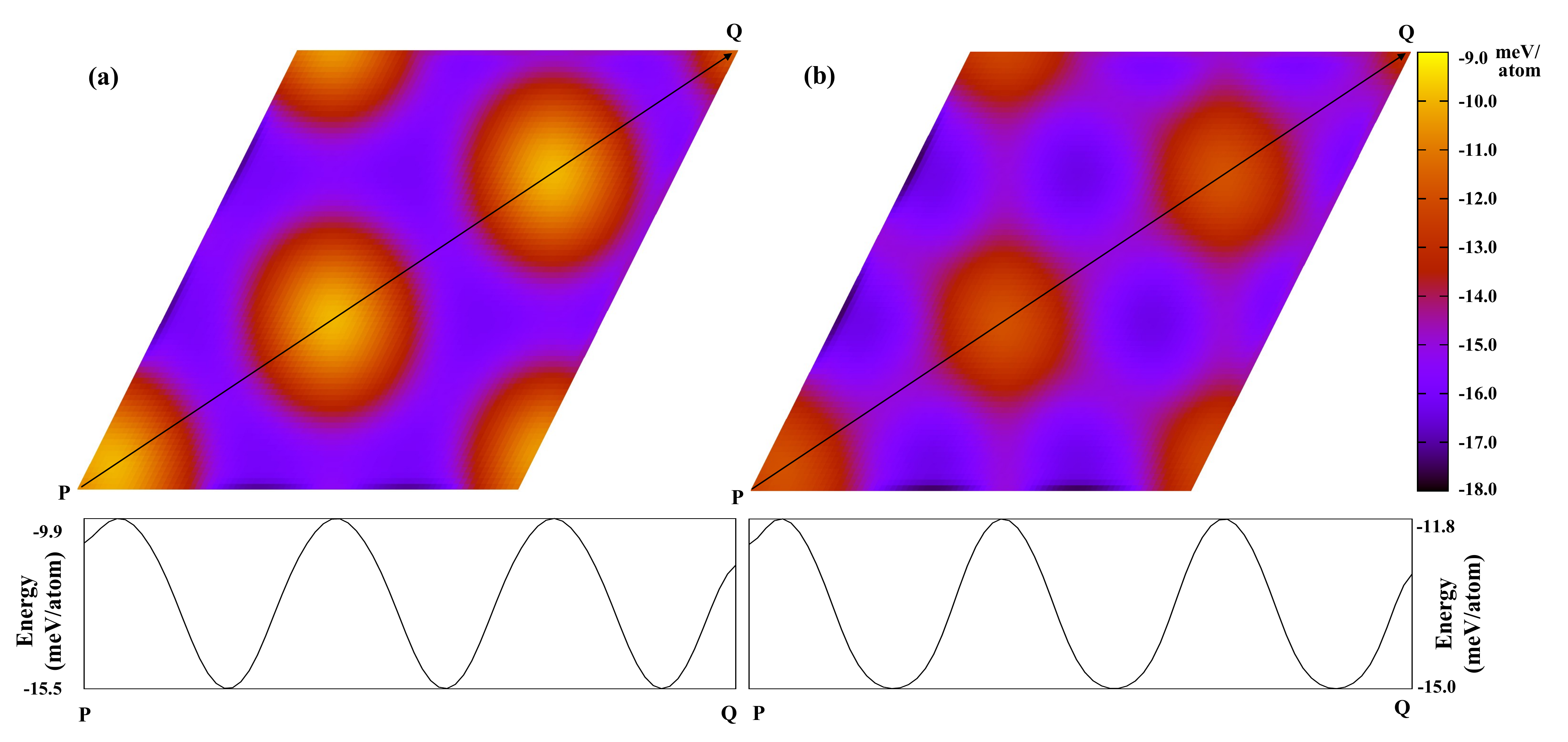}
\caption{Local energy profile of a sample having 15132 atoms with
$\theta=1.61^{\circ}$. (a) Before the relaxation. (b) After the
relaxation. The bottom panel depicts the local energy along the main diagonal
axis PQ which shows sinusoidal behaviour in this case.}
\label{figure_2}
\end{center}
\end{figure}

\begin{figure}
\begin{center}
\includegraphics[width=1.0\textwidth]{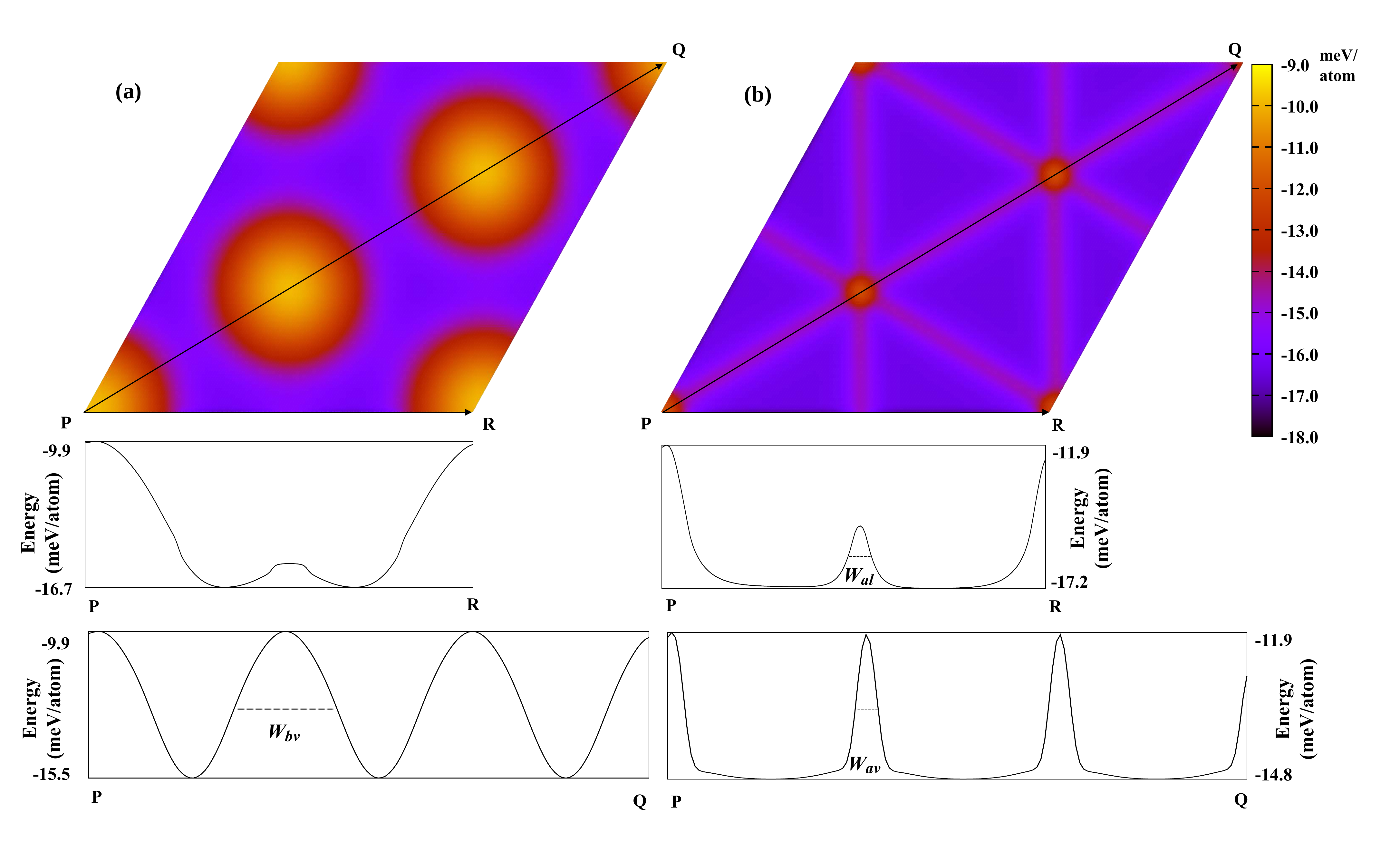}
\caption{Local energy profile of a sample having 321,492 atoms with
$\theta=0.35^{\circ}$. (a) Before the relaxation. (b) After the
relaxation. The bottom panels depict the local energy along the two principal
axes of the vortex lattice, horizontal PR and diagonal PQ. This shows
that sinusoidal behaviour is not present at smaller $\theta$ along the
PQ direction.}
\label{figure_3}
\end{center}
\end{figure}

\begin{figure}
\begin{center}
\includegraphics[width=1.0\textwidth]{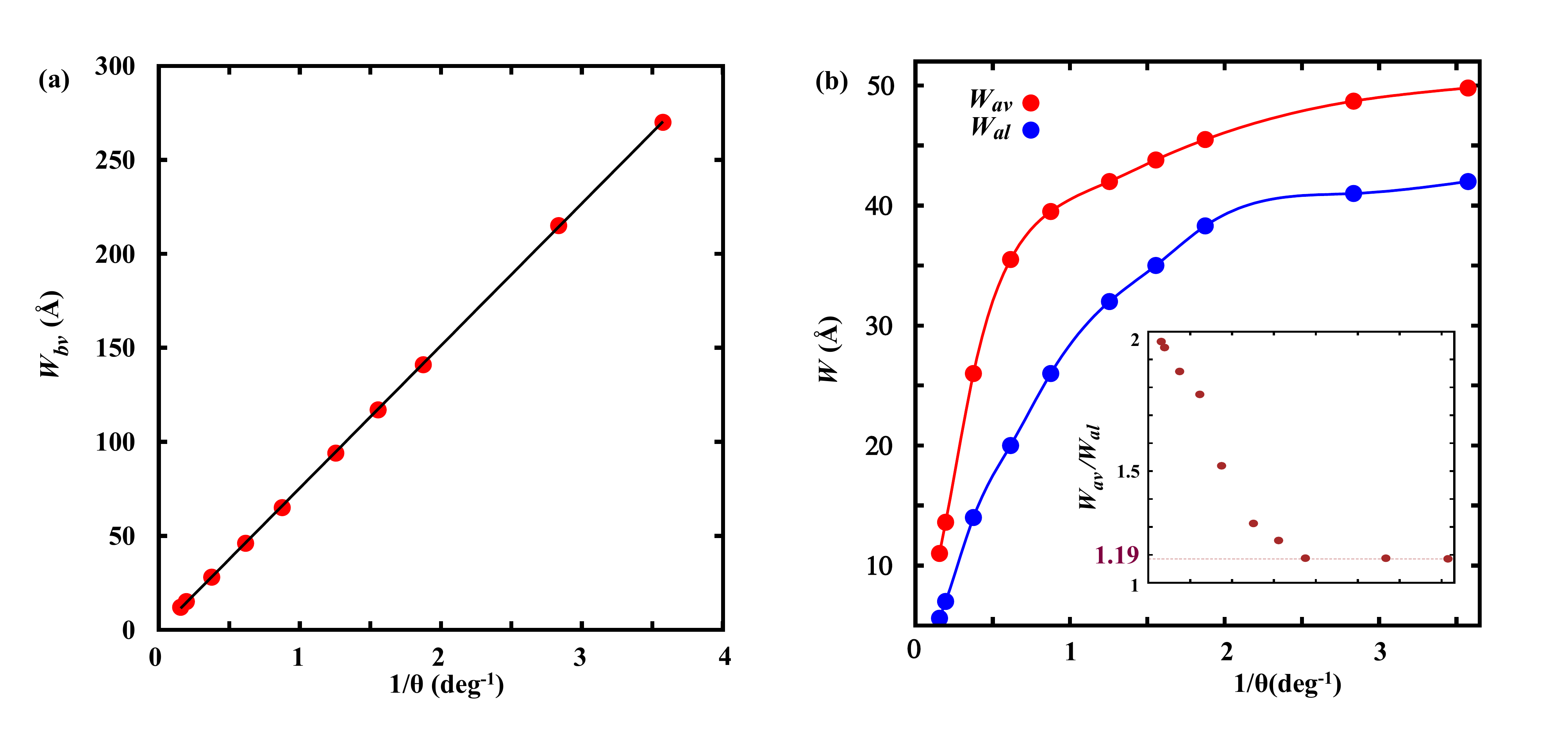}
\caption{Scaling behaviour of peak widths corresponding to the size of
the vortices as a function of inverse of the mismatch angle ($1/\theta$). We simulate the system sizes from 964
atoms ($\theta=6.40^{\circ}$) to  511,228 atoms ($\theta=0.28^{\circ}$).
(a) $W_{bv}$ (width of vortices peak at half its height, before relaxation) as a function of
inverse of the mismatch angle. We observe a linear scaling with $1/\theta$. Since $\theta$
scales as inverse of $L$, $W_{bv}$ scales linearly with system size. (b)
$W_{av}$ (width of vortices peak at half its height, after relaxation) and $W_{al}$ (width
of line peak at half its height, after relaxation) as a function of inverse of the mismatch angle. At large
small mismatch angle ($\theta<0.6^{\circ}$) the ratio between these two peak
widths becomes constant (inset).}
\label{figure_4}
\end{center}
\end{figure}

\begin{figure}
\begin{center}
\includegraphics[width=1.0\textwidth]{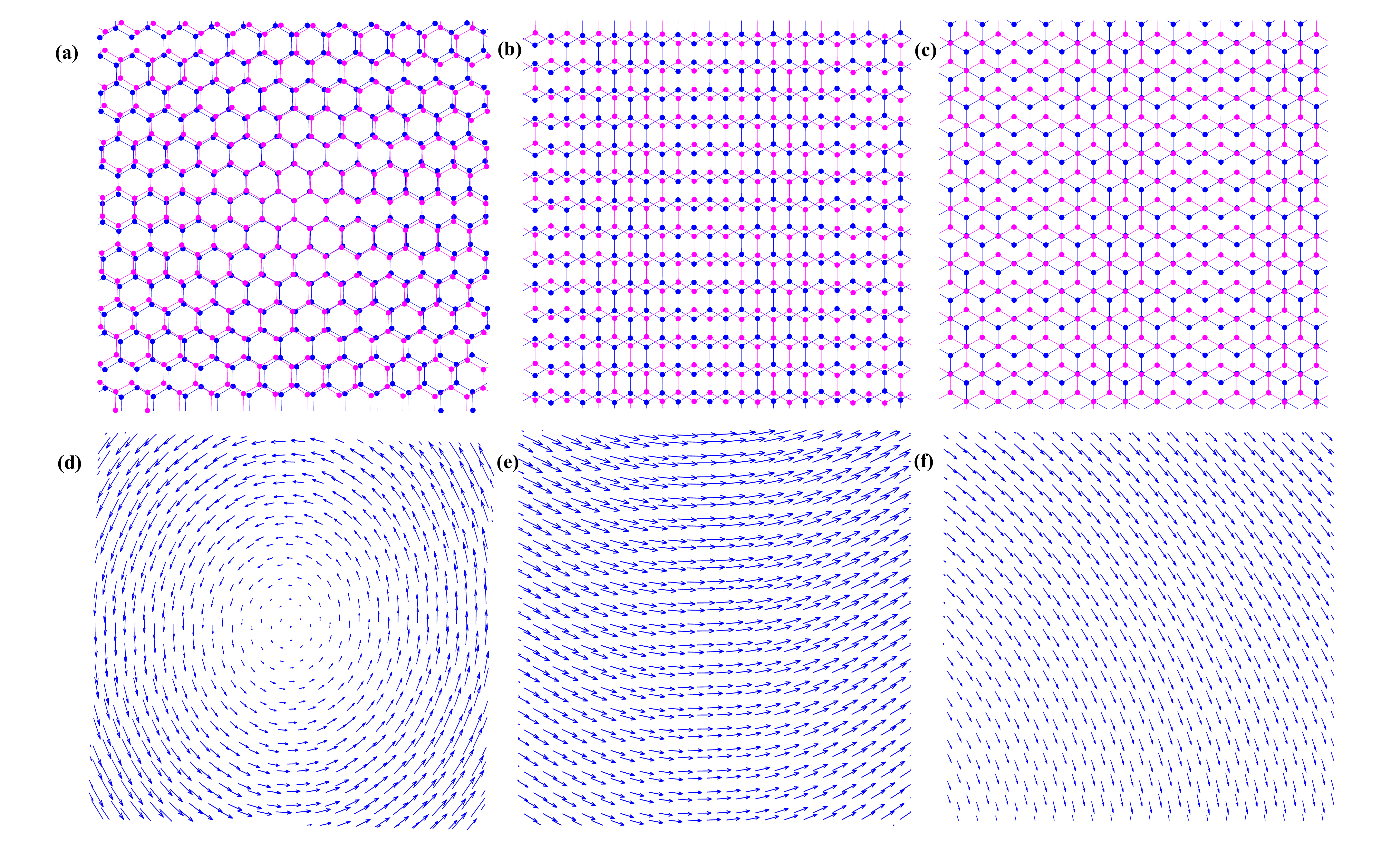}
\caption{Detailed structures and displacement fields around a vortex, line and Bernal stacked (AB/BA) region. (a-c) Atomic structures of a vortex, line and Bernal stacked (AB/BA) region, respectively. Here blue color is used for bottom layer and magenta color is used for top layer. (d-f) Displacement fields around a vortex, line and Bernel stacked (AB/BA) region in bottom layer with respect to their unrelaxed positions, respectively. }
\label{figure_5}
\end{center}
\end{figure}

To study the effect of relaxation qualitatively, we generate a sample
having 15132 atoms with a mismatch angle of $1.61^{\circ}$. The local energy
profile of the sample before and after the relaxation is shown in
Fig.\ \ref{figure_2}(a) and Fig.\ \ref{figure_2}(b), respectively. The
binding energy of AA and AB stacking after the minimization is 11.8
meV/atom and 17.5 meV/atom respectively which is in very good agreement
with reported values by Mostaani {\it et al.} calculated using quantum Monte
Carlo technique \cite{Mostaani2015}. The energy along the diagonal
principal axis PQ behaves as a sinusoidal function before and after
the relaxation for large values of the mismatch angle, as shown in
Fig.\ref{figure_2}. However, this sinusoidal behaviour of energy is no
longer present for small mismatch angles ($\theta<1^{\circ}$), as shown
in Fig.\ref{figure_3}. The elastic energy becomes rather concentrated at
the well-defined vortices in the displacement field. The local energy profile
of the sample having 321,492 atoms with mismatch angle of $0.35^{\circ}$,
before and after the relaxation is shown in Fig.\ref{figure_3}(a) and
Fig.\ref{figure_3}(b), respectively. Our simulations show that before
relaxation the size of the vortex around AA stacking increases linearly
with system size $L\sim\sqrt{N}\sim1/\theta$, with $N$ as the number
of atoms. The width of the peak at half maximum height along the diagonal PQ
before the relaxation is given as $W_{bv}$ and plotted as a function
of $1/\theta$ in Fig.\ref{figure_4}(a). Here, the subscript $b$ stands for
`before relaxation', and the subscript $v$ for `vortex'. Further on in this
manuscript, we will also use subscripts $a$ and $l$, which stand for
`after relaxation' and `line', respectively. We calculate the peak width
along the diagonal PQ after the minimization ($W_{av}$) and plot it
as a function of $1/\theta$ as shown in Fig.\ref{figure_4}(b). For
large system size $W_{av}$ appears to approach a constant value of $\sim$ $50$
\AA~. We also calculate the peak width after the minimization along the line PR,
represented as $W_{al}$. In the local energy profile of relaxed samples,
vortices are connected via a line which denotes a configuration with the
structure in-between AA and AB stacking as shown in Fig.\ref{figure_5}(b). The binding energy of this kind
of stacking is 14.8 meV.
We plot $W_{al}$ as a function of $1/\theta$ and find that at small mismatch angles (large system sizes) 
it also approaches a constant value of $\sim$ $42$
\AA~. The ratio between $W_{av}$ and $W_{al}$ becomes constant for all
the systems with mismatch angle below $0.6^{\circ}$ as shown in inset of
Fig.\ref{figure_4}(b). We find that the value for the constant ratio is $1.19$
in the thermodynamic limit.

The Bernal stacking in BLG has been investigated experimentally via STEM, where it has been shown that regions of AB and BA stacking are separated by nanometer wide rippled boundaries \cite{Butz2014,Lin2013}. In our simulations this is also the case as shown in Fig.\ref{figure_3} where lines connecting the vortices are separating AB and BA stackings. We present the detailed structures of these vortices, lines and Bernal stackings with displacement fields in Fig.\ref{figure_5}. Recent studies by Dai {\it et. al.} determining the size of the lines and vortices using the Peierls-Nabarro model \cite{Dai2016,Dai2016a} are in very good agreement (within 10\%) with our estimate of constant size in the thermodynamic limit. Alden {\it et. al.} use the Frenkel-Kontorava model \cite{Alden2013} and report a size which is significantly larger than experimental observation \cite{Butz2014}. 

With very small mismatch angles and thus very large Moir\'e patterns, most of the additional
energy, $\Delta E$, due to the Moir\'e pattern comes from the lines connecting the vortices,
as these lines grow with decreasing angle, while the vortices do not.
The additional energy due to the Moir\'e pattern is a combination of intralayer and
interlayer energy terms. The intralayer energy contribution decreases inversely proportional to the line width $w_{al}$,
while the interlayer energy contribution increases linearly with line width
\begin{equation}
\Delta E=a w_{al} L + b \frac{L}{w_{al}},
\end{equation}
where the parameter $a$ is determined by the energy difference between the different stackings, and $b$ is determined by the bulk modulus of a graphene layer.

In classical elastic bead spring models with a fixed extension, the extension per spring in the system decreases linearly with the number of the springs. With harmonic springs, the energy per spring scales quadratically with extension, and the total energy thus decreases linearly with the number of springs. Here in equation (1), $w_{al}$ is analogous to the number of the springs. Therefore, the intralayer energy contribution decreases inversely proportional to the line width $w_{al}$. The interlayer energy simply depends on the mismatched area in the sample and therefore scales linearly with $w_{al}$.

Minimizing $\Delta E$ with respect to the line width results in an $L-$independent $w_{al}$ given as
\begin{equation}
w_{al}=\sqrt{b/a}.
\end{equation}
Therefore, in the large samples where the size of vortices becomes constant, the width of the line connecting the vortices also becomes constant since it only depends on the bulk modulus of graphene and the type of stacking between two layers. In our numerical simulations we find the trend which is consistent with this analytical argument. We have calculated the value of interlayer energy constant as $a=0.0018$ eV/\AA$^{2}$ and intralayer energy constant as $b=3.1750$ eV by fitting our numerical energy data to equation (1). 

We minimize the samples in all directions for two different boundary conditions: deformation-free (DF) boundary conditions where the periodic box is determined by the crystalline structure of single undeformed graphene layers, and force-free (FF) boundary conditions where changes in the simulation box are allowed \cite{Jain2016}: the length of each of the periodicity vectors as well as the angle between them is determined by the constraint of minimal total energy. Our results on energetics of TBLG (Figs.\ 2-5) are based on DF boundary conditions since structures with DF boundaries allow us to compare atomic coordinates before and after relaxation without complications due to differences in box size. Moreover, we verified that the shrinkage in the box size and the differences in the energies between two different boundary conditions are very small ($ < 0.08 \%$) and do not alter the results and predictions presented in the paper. But this small decrease in the box size has very significant consequences on the buckling height, which we discuss next.

\begin{figure}
\begin{center}
\includegraphics[width=1.0\textwidth, height=20cm]{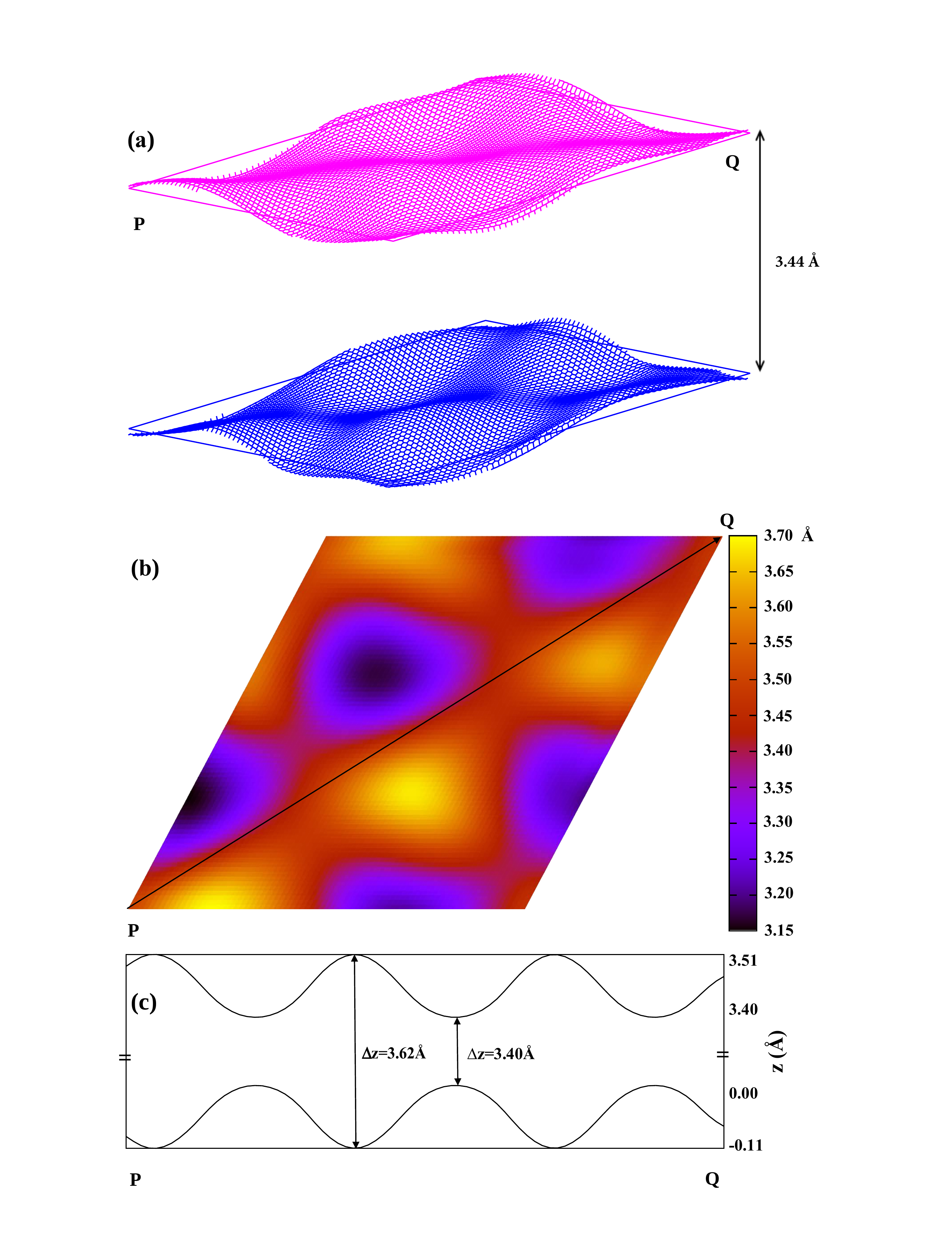}
\caption{Buckling behaviour in a sample with 15132 atoms
($\theta=1.61^{\circ}$). (a) Ripples in both top and bottom layer. The
equilibrium separation between both layers is $3.44$\AA~. (b) Buckling
profile of the top layer for DF boundary conditions. The
buckling height is $0.51$\AA~. (c) Buckling along the diagonal PQ in
both the layers. Around the AA stacked area the separation between the layers
is $3.62$\AA~.}
\label{figure_6}
\end{center}
\end{figure}

\begin{figure}
\begin{center}
\includegraphics[width=1.0\textwidth]{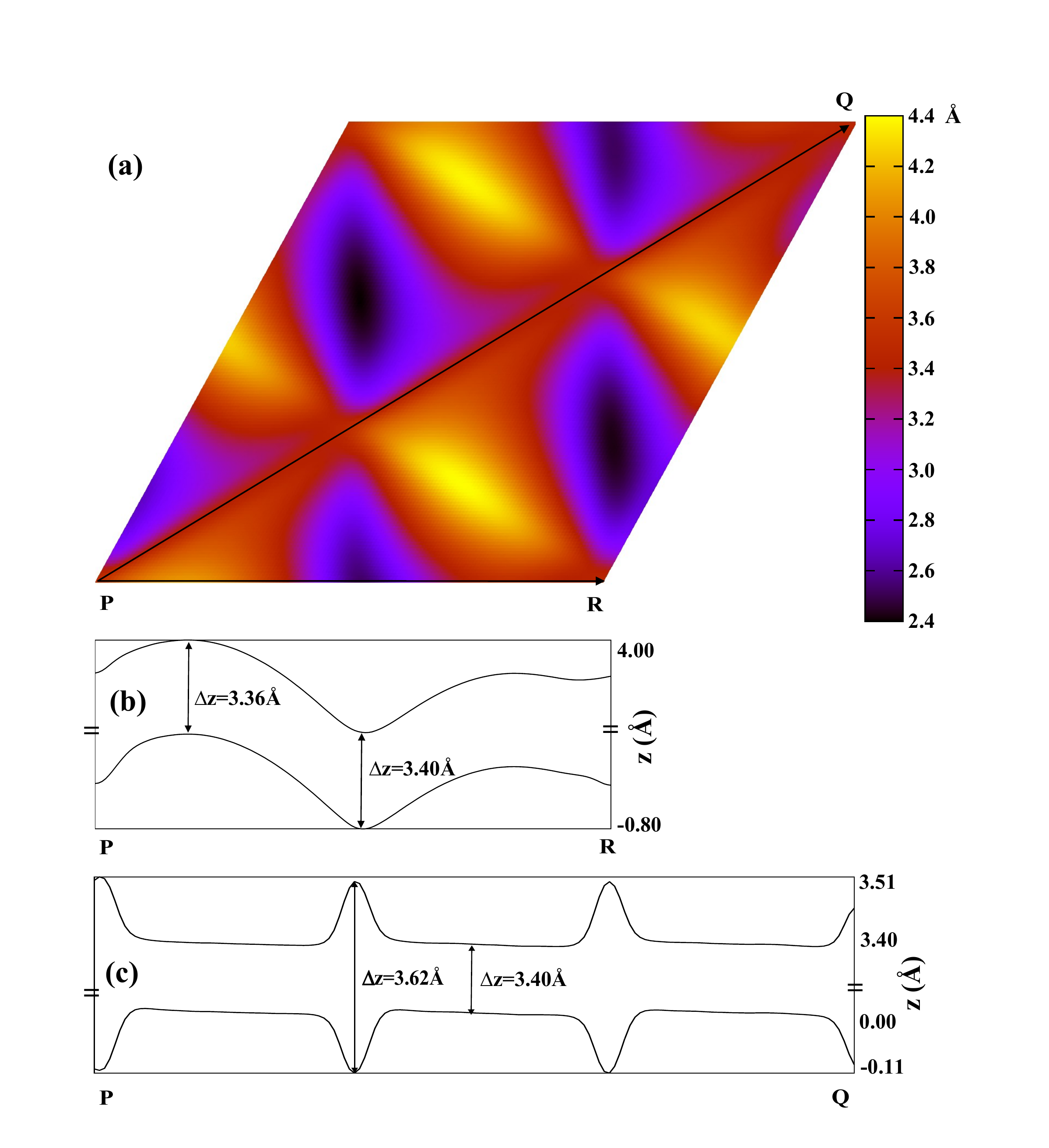}
\caption{Buckling behaviour in a sample with 321,492 atoms
($\theta=0.35^{\circ}$). (a) Buckling profile of the top layer. The
buckling height is $1.74$\AA~. (b) Buckling along the line PR in both
the layers. Around the AB stacked area the separation between the layers is
$3.36$\AA~. (c) Buckling along the diagonal PQ in both the layers. Around
the AA stacked area the separation between the layers is $3.62$\AA~. The
equilibrium separation between both layers is $3.38$\AA~. }
\label{figure_7}
\end{center}
\end{figure}

We now consider out-of-plane deformations in the TBLG samples. Our samples
before the relaxation have completely flat layers separated by $3.4$
\AA~in the $z$-direction. The minimized structures have out-of-plane
deformations characterized by the type of stacking between the
layers. In Fig.\ \ref{figure_6}(a) we show the structure of ripples
in a sample with $N=15132$ atoms after the complete relaxation. The
equilibrium average separation distance is $3.44$ \AA~ in between the
layers. The profile of out-of-plane deformations in the top layer is
shown in Fig.\ref{figure_6}(b). The buckling height in the individual
layer is $0.51$ \AA~ for DF boundaries. For FF boundaries the buckling
height is more significant and reaches a value of $1.12$ \AA~. The
out-of-plane deformations along the diagonal PQ direction are plotted
in both top and bottom layer, as shown in Fig.\ref{figure_6}(c). The
behaviour along the PQ direction is sinusoidal and the separation
around AA stacking is $3.62$ \AA~, in good agreement with previously
reported values in literature calculated using density functional theory
(DFT) calculations \cite{Uchida2014,Campanera2007}. Most importantly,
we observe a Moir\' e pattern-like feature in the buckling height,
see Figs.\ref{figure_6}(b) and \ref{figure_7}(a).

\begin{figure}
\begin{center}
\includegraphics[width=1.0\textwidth]{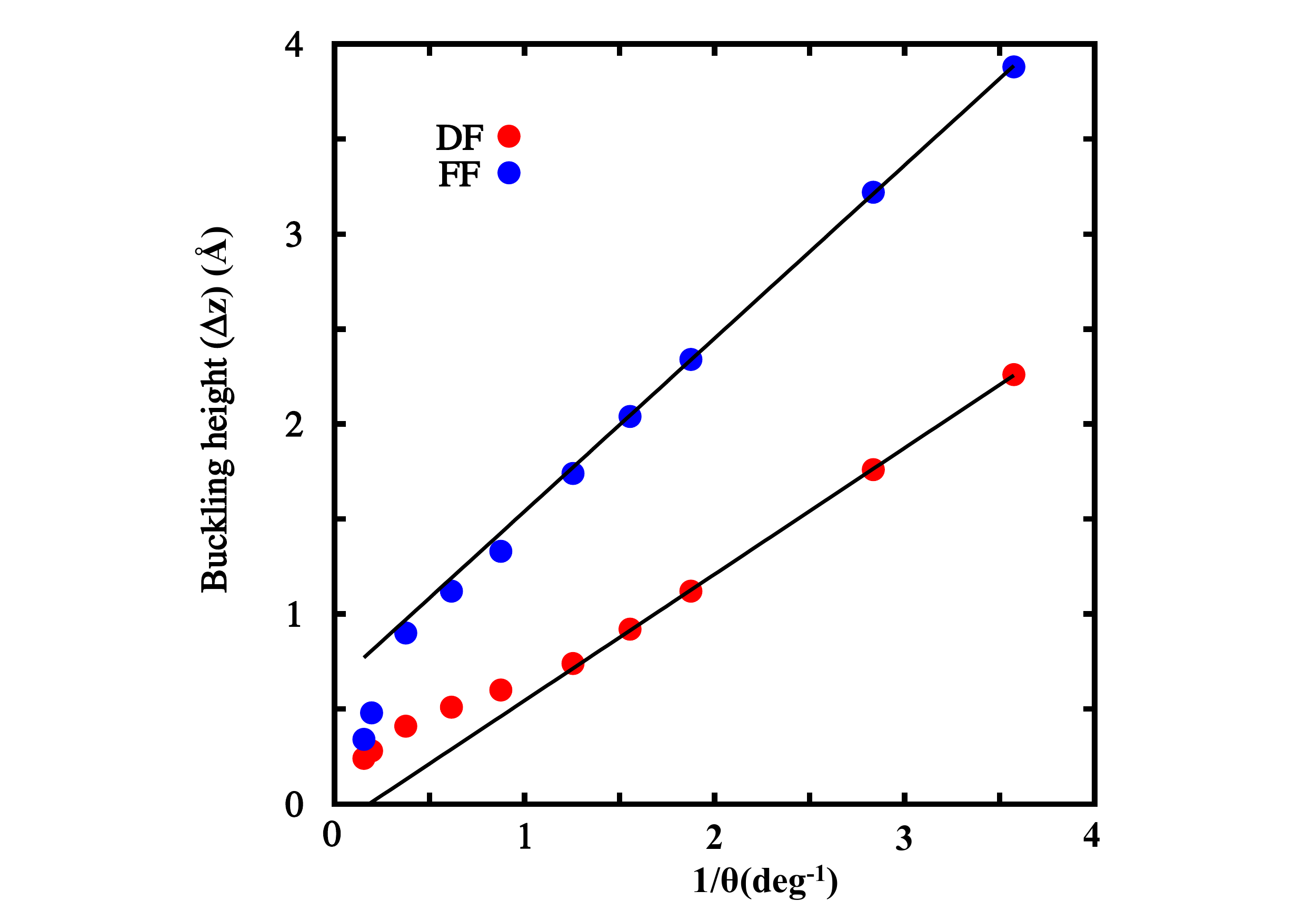}
\caption{Buckling height as a function of inverse of the mismatch angle ($1/\theta$) for both DF
and FF boundary conditions. For large system sizes (small $\theta$),the scaling of the
buckling height is linear with the system size. Intrinsic ripples are
significant and have the values of height $2.3$\AA~and $3.8$\AA~for DF
and FF boundary conditions, respectively, in a sample with $N=511,228$
atoms with $\theta=0.28^{\circ}$.}
\label{figure_8}
\end{center}
\end{figure}

We now discuss in more detail features of the spatial pattern in the
buckling height. As we already pointed out, with increasing system size the
vortex in the in-plane displacement around AA stacking shrinks after the
minimization and appears to become constant for $\theta<0.6^{\circ}$. This
feature yielding a characteristic length scale can also be seen in the
buckling of a sample having 321,492 atoms ($\theta=0.35^{\circ}$), as
shown in Fig.\ref{figure_7}. Namely, the characteristic length scale
in this case is the equilibrium average separation distance, and its
size relative to the system size decreases with increasing system size,
since the AB stacked area grows and AA stacked area does not. In this
case it has the value of $3.38$ \AA~. The sinusoidal behaviour in the
buckling, as shown in Figs.\ref{figure_6}(c), disappears for small
mismatch angles as shown in Figs.\ref{figure_7}(c) . Finally, the buckling height increases linearly
with system size for both DF and FF boundary conditions, as shown
in Fig.\ref{figure_8}. The buckling height for the largest sample
($N=511,228$ atoms with $\theta=0.28^{\circ}$) studied by our simulations
for FF boundary conditions is quite significant as the value is $3.78$ \AA~.

For the smallest twist angle under periodic boundary conditions ($\theta$ $\sim$ $1/L$ $\sim$ $1/\sqrt{N}$), each “mismatch line” seems to induce a small, constant buckling angle, which causes a buckling height that increases linearly with system size $L$. Without periodic boundaries, the twist angle is not discretized and can approach zero at any fixed system size; but we only simulated periodic boundaries. It is however clear that if at fixed $L$ the twist angle approaches zero, the buckling height has to approach zero as well, as the system then gradually approaches the perfectly aligned crystal, which is flat.

\section{CONCLUSIONS}
Our work demonstrates the crucial importance of having large, well-relaxed
samples of twisted bilayer graphene, to study its structural properties. The new
combination of intralayer and interlayer potentials uses explicit
lists of bonds and is therefore computationally very cheap. This
allows us to accurately simulate very large TBLG samples with very small mismatch
angles. The simulation results are in very good agreement with reported in the
literature. There are sinusoidal modulations in the energy and buckling height for large
misorientation angles but this behaviour no longer persists at small
misorientation angles. We have shown with large scale atomistic simulations that
the size of the vortices in the displacement field approaches a constant
in the thermodynamic limit. There are significant out-of-plane deformations
which increase with increasing system size. The characteristic average
separation between the layers also becomes constant in the thermodynamic
limit. These structural properties should have direct effect on electronic
and optical properties of twisted bilayer graphene. In future work, the same
combination of potentials can be modified with different structural
parameters to investigate other misalinged two-dimensional materials
such as h-BN, MoS$_{2}$ and WSe$_{2}$.

\ack
This project is funded by FOM-SHELL-CSER programme (12CSER049). This
work is part of the research programme of the Foundation for Fundamental
Research on Matter (FOM), which is part of the Netherlands Organisation
for Scientific Research (NWO).

\section*{References}
\bibliographystyle{iopart-num}
\bibliography{Bi-graphene}
%\begin{thebibliography}{}

%\end{thebibliography}

\end{document}